# Analyzing Economic Convergence Across the Americas: A Survival Analysis Approach to GDP per Capita Trajectories.

**Diego Vallarino**
Independent Researcher
Madrid, Spain
diego.vallarino@gmail.com

April 2024

**Abstract:**

By integrating survival analysis, machine learning algorithms, and economic interpretation, this research examines the temporal dynamics associated with attaining a 5 percent rise in purchasing power parity-adjusted GDP per capita over a period of 120 months (2013-2022). A comparative investigation reveals that DeepSurv is proficient at capturing non-linear interactions, although standard models exhibit comparable performance under certain circumstances.

The weight matrix evaluates the economic ramifications of vulnerabilities, risks, and capacities. In order to meet the GDPpc objective, the findings emphasize the need of a balanced approach to risk-taking, strategic vulnerability reduction, and investment in governmental capacities and social cohesiveness. Policy guidelines promote individualized approaches that take into account the complex dynamics at play while making decisions.

## 1. Introduction

In contemporary economic research, the exploration of temporal dynamics in a nation's journey to achieve a specific level of GDP per capita gains paramount importance. This empirical investigation, conducted across 33 American countries, adopts a nuanced approach by incorporating a comprehensive dataset that includes countries with right-censored data (9 countries) and those reaching a 5% increase in GDP per capita at purchasing power parity (PIBpcPPP) within 120 months (24 countries).

In addressing the central query, this research aims to unravel the intricate relationship of variables and risks influencing the time required for a country to achieve the specified 5% increase in GDP per capita. Leveraging advanced statistical techniques, particularly survival analysis, the study incorporates key variables such as Vul_Inherent, Vul_Fragility_Democracy, and Vul_Human Rights, offering a robust understanding of multifaceted vulnerabilities.

This academic pursuit emphasizes rigorous methodologies, empirical analyses, and data-driven insights. With a focus on a time-centric lens, the study contributes not only to theoretical advancements in economic literature but also to the formulation of more effective policies aimed at expediting and sustaining economic growth. The introduction of models commonly used to study GDP, such as the neoclassical growth model, the Keynesian model, the Solow growth model, and the Harrod-Domar model, lays the foundation for a comprehensive exploration (Dutta & Mishra, 2023; Nolan et al., 2019; Yunita et al., 2023).

Moving forward, the research addresses the apparent gap in survival analysis within GDP per capita prediction specialists. It evaluates the effectiveness of various machine learning survival models, including Cox, Kernel SVM, DeepSurv, Survival Random Forest, and MTLR models, using the concordance index to compare their predictive capabilities. The research's primary goal is to determine the most accurate model for projecting the time until a country reaches a 5% increase in GDP per capita at purchasing power parity (PIBpcPPP) within 120 months.

To achieve this goal, the study scrutinizes the significance and magnitude of vulnerabilities and risks associated with the Multidimensional Vulnerability Index (MVI) of the United Nations Development Programme, comparing the results to economic theory and intuition (Assa, J., et al., 2023; OAS, 2023). The research seeks insights into the factors influencing GDP per capita levels and aims to provide a better understanding of how survival models can inform public policies.

The paper's structure includes a theoretical perspective on survival analysis models, a brief examination of works using survival analysis to address financial collapse, an empirical analysis encompassing models, data sources, and assessment measures, analytical results, and a conclusion summarizing key findings and implications for future research and policymaking. Overall, this research contributes to the literature on survival analysis in finance, offering insights into features leading to financial catastrophes and aiding policymakers in developing effective regulations to prevent such disasters in the future.

## 2. Theoretical perspective

Survival analysis is a crucial instrument in the field of economic analysis, providing a nuanced perspective for scrutinizing dynamic processes like the lifespan of businesses, the effects of economic policies, and the enduring nature of economic events. Survival analysis has considerable importance within the realm of economics due to its capacity to include time-to-event data, so furnishing a resilient framework for evaluating the durations till certain economic occurrences transpire (Jin et al., 2021; P. Wang et al., 2017; Zelenkov, 2020; Zhou et al., 2022). Nevertheless, its efficacy is contingent upon a thorough comprehension of the characteristics and caliber of the accessible data.

It is critical to identify the nature of the data being analyzed, since errors or misunderstandings may significantly compromise the dependability of survival projections. In the realm of academic research, when confronted with the convergence of survival analysis and economic investigation, it is crucial to engage in a rigorous assessment and discernment of the complexities inherent in the data environment. This entails guaranteeing that the analyses performed not only enhance our comprehension of temporal economic phenomena but also provide the foundation for more informed decision-making processes (Finch, 2005; Gorfine & Zucker, 2022; Maharana et al., 2022; Mumuni & Mumuni, 2022).

Several distinct forms of data may have an impact on the use of these strategies. For example, fundamental patient data may include demographic and clinical particulars, including the stage of illness, concurrent conditions, and course of therapy (Hair & Fávero, 2019; Maharana et al., 2022).

The use of datasets including censored data, competitive risk data, or longitudinal patient information might possibly give rise to complications (Barrett et al., 2011). Survival algorithms remain relevant notwithstanding some limitations (Cuperlovic-Culf, 2018; Jin et al., 2021).

We will address the significance of this kind of data for survival analysis in the following paragraphs.

*2.1 Baseline Agent Data*

Fundamental patient information is critical for healthcare professionals to formulate a survival plan. Demographic information, including race, ethnicity, age, and gender, may have a significant impact on a patient's survival rate, in conjunction with clinical data like disease stage, comorbidities, and treatment history (Hair & Fávero, 2019; Haradal et al., 2018; Mumuni & Mumuni, 2022). While we have illustrated the attributes of the baseline data using the most representative instance from the survival analysis literature, the majority of these attributes are applicable to the examination of customers during the purchasing phase, employees throughout the work process within the organization, or the company throughout its lifespan.

Constructing a survival plan requires comprehensive comprehension and evaluation of several aspects that might potentially impact the prognosis of a patient (Cuperlovic-Culf, 2018; Jin et al., 2021). Healthcare practitioners may provide survival estimates using a variety of methods, such as forests, trees, neural networks, deep learning, multitasking, boosting, and "extras" (Thenmozhi et al., 2019; Zhao et al., 2022). In order to mitigate errors and biases, it is important to take into account the limitations and constraints of these algorithms while producing forecasts (Azodi et al., 2020). Hence, although core patient data is essential for developing precise survival algorithms that ensure effective patient care, it is often inadequate to attain a decent level of performance in machine learning models.

*2.2 Censored Data*

Survival data is identifiable by the concept of data suppression. In the case when the event under investigation is the demise or insolvency of a corporation, the event time for surviving participants is suppressed at the completion of the study. This necessitates the continuation of the statistical analysis in the absence of the subject's date of death (Basak et al., 2022a; Jiang, 2022; Vinzamuri et al., n.d.).

The only accessible information about his demise is that it happened subsequent to the conclusion of the investigation. People who discontinue follow-up research are often subject to censorship, since their occurrence is frequently overlooked and their timing is uncertain (Raghunathan, 2004). The unobserved date of the event does not qualify as a missing data point, since these two classifications of unobserved data possess unique characteristics and are subject to different empirical interpretations (Yuan et al., 2022).

The only known information about right-censored subjects is that the occurrence in question occurred subsequent to the censorship era. Had the study been sustained (or had the volunteers remained), the outcome of interest to all participants would have been seen in the end (Basak et al., 2022a, 2022b; Jiang, 2022). Conventional statistical methods for examining survival data operate under the assumption that filtering is non-informative or independent (Khan & Zubek, 2008).

This means that, at a specific juncture, the subjects who are still under follow-up have an equivalent future risk of experiencing the event as those who are no longer under follow-up (either because of censorship or study abandonment). This would be the case if the losses to follow-up were arbitrary and thus lacking in informative value (Basak et al., 2022b).

Present study unequivocally establishes that proper management of censored data is critical for obtaining an accurate depiction of the upcoming survival analysis experiment (Jiang, 2022). Consequently, this research will aim to ascertain the most effective methods for including censored data, encompassing both the left and the right (which is the predominant approach in analytic models). While the literature has not extensively examined the latter, time-to-event statistical analysis may provide valuable insights (Cui et al., 2020; Yuan et al., 2022).

Censored data is a common occurrence when working with survival data; it occurs when the exact date of an event is unclear, but it is evident that the event did not transpire before to or subsequent to a certain time period. Left-censored data, interval-censored data, and right-censored data are the three types of censored data. A multitude of exceptional methods are now accessible to handle enormous volumes of filtered data (Cui et al., 2020; Yuan et al., 2022).

Survival Random Forest is one approach that can handle constrained data well. It is a machine learning technique that integrates the predictions of several decision trees (Jin et al., 2021; Jin Ziweiand Shang, 2020; Zhao et al., 2022). Multi-Tasking Linear Regression (MTLR) is an extra method for managing censored data in an efficient manner. It estimates the survival time distribution using a Bayesian method, which is advantageous when dealing with several outcomes (L. Wang et al., 2017). XGboost, an additional well-known algorithm, is capable of processing vast quantities of censored data including categorical and continuous variables (Barnwal et al., 2022).

Subsequent portions of this study will undertake an investigation into the predictive performance of distinct machine learning models for survival analysis with regard to the time required for various nations to attain a 5% increase in GDP per capita at purchasing power parity (PIBpcPPP) within 120 months. The objective of this inquiry is to critically assess the effectiveness and relative merits of various models in representing the temporal intricacies that are intrinsic to economic growth. Through an assessment of the prognostic capacities of machine learning algorithms within this particular framework, our objective is to provide contributions that enhance our comprehension of the temporal aspects linked to the achievement of particular levels of GDP per capita. Such insights would have significant ramifications for the formulation of policies and the prediction of economic conditions.

## 3. Empirical Analysis

This empirical investigation intricately examines the dynamics of economic growth across 33 countries, leveraging a nuanced approach to create a comprehensive dataset. The dataset encompasses countries with right-censored data (9) and those achieving a 5% increase in GDP per capita at purchasing power parity (PIBpcPPP) within 120 months (24).

This meticulous sampling strategy ensures a thorough representation of diverse economic scenarios, incorporating information from authoritative sources such as the World Bank, IMF, and the United Nations (Assa et al., 2023; OAS, 2023).

With the intention of answering the primary inquiry, this study endeavors to decipher the complex interconnections among factors and hazards that impact the duration needed for a nation to attain the designated 5% augmentation in GDP per capita at purchasing power parity. By using sophisticated statistical methods, including survival analysis, this research aims to provide a comprehensive comprehension of the temporal dimensions of economic growth.

The analysis incorporates key variables, including Vul_Inherent, Vul_Fragility_Democracy, and Vul_Human Rights, which have been meticulously curated to capture multifaceted vulnerabilities.

### 3.1 Models

*3.1.1 Cox Proportional Hazards Model (coxph)*

The Cox proportional hazards model is a widely used semi-parametric model in survival analysis. It assumes that the hazard function can be represented as the product of a time-independent baseline hazard function and a time-varying covariate function. Mathematically, the model can be represented as:

$$h(t|x) = h_0(t) \exp(\beta^T x)$$

where $h(t|x)$ is the hazard function for a given time t and covariate values $x$, $h0(t)$ is the baseline hazard function, $\beta$ is a vector of regression coefficients, and $exp(\beta X)$ is the hazard ratio, which represents the change in hazard associated with a unit change in the covariate.

*3.1.2 Multi-Task Logistic Regression (MTLR)*

Multi-task logistic regression is a machine learning method that can be used for survival analysis. It is a multi-output learning algorithm that can predict the probability of an event occurring at different time points. Mathematically, the model can be represented as:

$$h(t|x) = exp\left(\Sigma_{k=1}^{K} \Sigma_{j=1}^{p} \beta_{kj} x_{kj}\right)$$

Where $h(t|x)$ is the hazard rate for an individual with covariates x, $\beta_{kj}$ are the regression coefficients for the kth characteristic of the jth group, and $x_{kj}$ is the kth feature of the jth group.

*3.1.3 Kernel Support Vector Machine (Kernel SVM)*

Kernel support vector machines are a popular machine learning method for survival analysis. They can handle non-linear relationships between covariates and outcomes by projecting the data into a higher-dimensional space using a kernel function. The model can be represented as:

$$f(x) = sign(\Sigma_{i=1}^{n} \alpha_i y_i K(x_i, x) + b)$$

Where $K(x_i, x)$ is a kernel function that measures the similarity between the feature vectors $x_i$ and $x$, $y_i$ is the class label of the i-th instance, $\alpha_i$ are the weights of the support vectors and $b$ is the bias.

*3.1.4 Random Survival Forest*

Random survival forests are an extension of random forests for survival analysis. They use an ensemble of decision trees to predict the survival function. The model can be represented as:

$$h(t|x) = (1/B)\Sigma_{b=1}^{B} h_b(t|x)$$

Where $h_b(t|x)$ is the hazard rate for an individual with covariates $x$ in the $bth$ decision tree and $B$ is the number of trees in the random forest.

*3.1.5 DeepSurv*

DeepSurv is a deep learning model for survival analysis. It uses a neural network with a flexible architecture to predict the survival function. The model can be represented as:

$$h(t|x) = exp\left(\Sigma_{i=1}^{p} \beta_i f_i(x) + g(h_\theta(x))\right)$$

Where $h(t|x)$ is the hazard rate for an individual with covariates $x$ and $\beta_i$ are the regression coefficients for the input features $f_i(x)$, $g(\cdot)$ is a non-linear function that transforms the output features and $h_\theta(x)$ is a neural network with θ parameters.

**3.2 Data**

This empirical study looks at the patterns of economic development in 33 nations from all America (Cuba and Venezuela have been left out of the analysis because they do not have all the necessary data). Using a sophisticated methodology, the dataset includes nations that have been right-censored (9), and those that have reached the event within 120 months (24). This extensive sampling technique guarantees a complete representation of many economic situations.

For the analysis, the experimental setup now considers the event as the country achieving a 5% increase in per capita GDP at purchasing power parity (PIBpcPPP). The observation spans from 2013 to 2022. If the 5% threshold has not been reached within these 120 months, the *status* is designated as 0. Conversely, if the country has achieved the specified per capita GDP level, the status is set to 1. Notably, adjustments have been made to mitigate the impact of changes in 2020 due to the COVID-19 pandemic. Furthermore, the study incorporates information from authorized sources such as the World Bank, IMF, and UN, as well as data from 120 nations to identify multiple risks (see Annex).

The dataset is a comprehensive compilation of various variables, each offering a unique perspective on economic scenarios (Balica et al., 2023; Dutta & Mishra, 2023). The primary variables include:

**Natural_Risk:** EM-DAT characterizes disasters as occurrences or situations that surpass local capacity, compelling the need for external assistance at the national or international level. These events are typically unforeseen and abrupt, resulting in substantial damage, destruction, and human suffering.

**Commercial_risk:** This variable captures the commercial risks associated with countries and is instrumental in understanding the economic challenges related to trade and market conditions.

**Financial_risk:** Focused on economic stability, Financial_risk encompasses indicators like the Emerging Markets Bond Index (EMBI) and Risk Rating S&P, providing insights into financial vulnerabilities.

**Endogenous_risk:** Examining internal economic factors, Endogenous_risk incorporates annual GDP growth, current account balance, inflation, primary balance, public debt, and external debt, offering a holistic view of a country's economic health.

**Vul_Inherent:** A composite variable, Vul_Inherent encapsulates critical aspects such as proximity to global markets, landlocked status, coastal population proportion, inhabitants in arid lands, total economic loss, fatalities, and affected individuals, providing a comprehensive measure of inherent vulnerabilities.

**Vul_Fragility_Democracy:** Focusing on democratic institutions, this variable includes indicators like expanded freedom, freedom of association, clean elections, suffrage share, and an elected official's index, offering insights into the fragility of democratic structures. Expanded freedom, freedom of association, clean elections, share of population with suffrage, elected officials index.

**Vul_Human Rights:** This variable focuses on equal treatment and the absence of discrimination. It assesses the effective guarantee of the right to life and security of the person, due process of the law, rights of the accused, freedom of opinion and expression, freedom of belief and religion, freedom from arbitrary interference with privacy, freedom of assembly and association, and fundamental labor rights.

**Vul_Homes:** Capturing household-level vulnerabilities, this variable includes the Human Development Index, multidimensional poverty, gender inequality incidence, Gini coefficient, and personal remittances, reflecting the socio-economic conditions at the household level.

**Vul_Companies:** Assessing business-related challenges, Vul_Companies includes indicators like ease of doing business, permits for construction, property registration, credit accessibility, international trade, and contract enforcement, offering a nuanced perspective on the business environment.

**Capabilities_State:** Focused on governance and infrastructure, this variable incorporates indices such as the corruption perception index, government effectiveness, Hyogo framework, access to electricity, internet users, adult literacy rate, cell phone subscriptions, road length, basic water services, basic sanitation, doctor density, MCV2 vaccine coverage, DTP3 vaccine coverage,

PCV23 vaccine coverage, national health expenditure per capita, and maternal mortality, providing insights into the state's capacity and performance.

**Social_Cohesion_Capabilities:** Exploring societal dynamics, this variable encompasses power distance, individualism, masculinity, uncertainty avoidance, long-term orientation, indulgence, civil society participation index, direct popular vote index, local government index, and regional government index, shedding light on social cohesion and governance.

**Time**: The temporal dimension signifies the duration from 2013 to 2022, capturing the span of a country to achive a 5% increase in GDP per capita at purchasing power parity (PIBpcPPP). The temporal dynamics during this period play a pivotal role in shaping the strategies and decisions of countries in their economic development.

**Status**: The status variable reflects the present condition of a country within the analysis period. It is binary, taking values of 0 or 1, signifying whether the country has reached or surpassed the 5% increase in GDP per capita at purchasing power parity (PIBpcPPP). A value of 0 indicates that the country has not achieved the 5% threshold, while a value of 1 signifies that the country has attained or surpassed this increment. This status variable provides a snapshot of the country's economic performance and its alignment with the specified GDP per capita increase, distinguishing between countries that persist in their economic development and those that may face challenges or have ceased to exist in the market.

### 3.3 Metrics

*3.3.1 C-Index*

The C-index (also known as the concordance index or the area under the receiver operating characteristic curve) is a widely used metric in survival analysis and medical research to assess the performance of predictive models that estimate the likelihood of an event occurring over a given time period.

The C-index is generated using the rankings of anticipated event occurrence probability for each participant in a dataset. It calculates the percentage of pairings of people in whom the person with the higher anticipated probability experienced the event before the person with the lower projected probability. In other words, it assesses a predictive model's capacity to rank people in order of their likelihood of experiencing the event of interest.

The C-index scales from 0 to 1, with 0.5 representing random prediction and 1 indicating perfect prediction. In medical research, a C-index value of 0.7 or above is considered satisfactory performance for a prediction model. Here is the formula of censored data C-Index.

$$C-index = \frac{\Sigma_{ij}\ 1_{T_j<T_i} \cdot 1_{\eta_j>\eta_i} \cdot \delta_j}{\Sigma_{ij}\ 1_{T_j<T_i} \cdot \delta_j}$$

$\eta_i$, the risk score of a unit $i$
$1_{T_j<T_i} = 0\ \ if\ T_j < T_i\ else\ 0$
$1_{\eta_j<\eta_i} = 0\ \ if\ \eta_j < \eta_i\ else\ 0$
$\delta_j$, represents whether the value is censored or not

## 4. Results

The regression analysis examines the Multidimensional Vulnerability Index (MVI) as the dependent variable, incorporating various independent variables representing distinct aspects of risk and vulnerability. The interpretation of economically significant results is outlined below.

The intercept (0.169364) represents the estimated MVI when all other independent variables are zero, indicating inherent vulnerability in the absence of specific risks or factors. Each coefficient for commercial risk, financial risk, endogenous risk, and other variables signifies the change in MVI associated with a one-unit increase in the respective independent variable, holding all others constant (see Annex).

Upon closer examination of the coefficients, *Natural_risk*: The coefficient of 0.126599 is statistically significant (p-value = 6.65e-07), suggesting that an increase in natural risk corresponds to an elevated level of multidimensional vulnerability. *Commercial_risk*: The coefficient of -0.047992 is not statistically significant (p-value = 0.325703), indicating that there is insufficient evidence to conclude a significant impact of commercial risk on the MVI.

*Financial_risk*: The coefficient of 0.070071 is statistically significant (p-value = 0.004069), implying that heightened financial risk is linked to increased vulnerability. *Endogenous_risk*: The coefficient of 0.047028 is not statistically significant (p-value = 0.358206), suggesting a non-significant positive impact on the MVI. *Vul_Inherent*: The coefficient of 0.000374 is not statistically significant (p-value = 0.994192), implying that inherent vulnerabilities alone do not have a significant impact on the MVI.

*Vul_Companies*: The coefficient of 0.128304 is statistically significant (p-value = 0.003753), indicating that vulnerabilities in companies contribute to a higher level of multidimensional vulnerability. *Vul_Homes*: The coefficient of 0.157584 is statistically significant (p-value = 0.000229), suggesting that vulnerabilities in households contribute significantly to the MVI. *Capabilities_State*: The coefficient of 0.004730 is not statistically significant (p-value = 0.932943), indicating a non-significant impact of state capabilities on the MVI.

*Social_Cohesion_Capabilities*: The coefficient of 0.167742 is statistically significant (p-value = 6.33e-05), suggesting that social cohesion in terms of capabilities significantly contributes to a higher level of multidimensional vulnerability.

Model statistics reveal a high multiple coefficient of determination (R-squared) of 0.9617, indicating that approximately 96.2% of the variability in the MVI is explained by the included independent variables (see Annex).

As we delve further into survival analysis, the Kaplan-Meier curve visually depicts the survival probabilities of a cohort of 33 countries over the 120-month duration. The x-axis signifies the time taken to reach a 5% increase in GDP per capita at purchasing power parity (PIBpcPPP). The y-axis portrays the survival probability, where all countries start with a survival probability of 1, indicating they are 'alive' at the beginning. Over time, certain countries may encounter failure, resulting in a gradual decline in survival probability.

**Figure 1: Kaplan-Meier survival curve.**

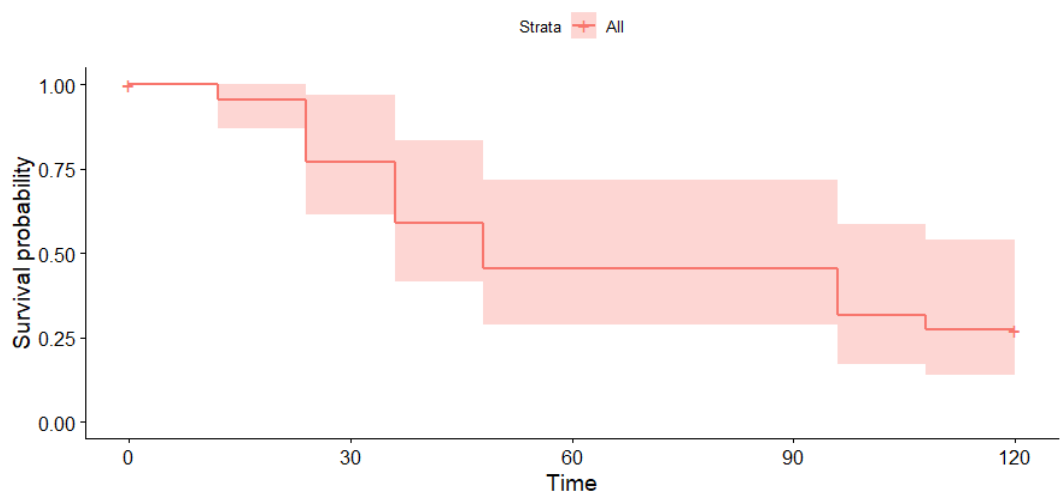

**Source**: own elaboration

The Table 1 supplies a comprehensive breakdown of the evolution of country failure risk at specific time points based on the survival analysis. For instance, at 10 months, 22 countries were at risk, and there were no events (failures), resulting in a survival probability of 1.000. This indicates that all countries were expected to reach a 5% increase in GDP per capita at purchasing power parity (PIBpcPPP) up to that point, reflecting a 100% survival probability.

As time progresses, the number of countries at risk decreases, accompanied by an increase in events (failures), leading to a gradual reduction in survival probabilities. At 50 months, 10 countries were at risk, with 12 events, resulting in a survival probability of 0.455. This suggests that around 45.5% of countries were expected not to reach a 5% increase in GDP per capita at purchasing power parity (PIBpcPPP) up to that point.

**Table 1: Kaplan-Meier survival probabilities (survival) at different time points**

| Call: surfit(formula = Surv(time, status) ~ 1, data = data.train | | | | | | |
|---|---|---|---|---|---|---|
| time | n.risk | n.event | survival | std.error | lower 95% CI | upper 95% CI |
| **10** | 22 | 0 | 1.000 | 0.0000 | 1.000 | 1.000 |
| **50** | 10 | 12 | 0.455 | 0.1062 | 0.288 | 0.718 |
| **80** | 10 | 0 | 0.455 | 0.1062 | 0.288 | 0.718 |
| **105** | 7 | 3 | 0.318 | 0.0993 | 0.173 | 0.587 |
| **108** | 7 | 1 | 0.273 | 0.0950 | 0.138 | 0.540 |
| **111** | 6 | 0 | 0.273 | 0.0950 | 0.138 | 0.540 |

**Source**: own elaboration

Continued analysis reveals a consistent decrease in survival probabilities. For example, at 80 months, 10 countries were at risk, but there were no events, resulting in a survival probability of 0.455. This indicates that approximately 45.5% of countries were expected not to reach a 5% increase in GDP per capita at purchasing power parity (PIBpcPPP) up to that point.

Further examination at 105 months shows that 7 countries were at risk, with 3 events, resulting in a survival probability of 0.318. This implies that about 31.8% of countries were expected not to reach a 5% increase in GDP per capita at purchasing power parity (PIBpcPPP) up to that point. The subsequent time points, at 108 and 111 months, demonstrate a consistent survival probability of 0.273, indicating that approximately 27.3% of countries were expected not to reach the median GDP per capita during these periods.

These findings underscore the dynamic nature of the risk landscape for countries, with decreasing survival probabilities suggesting heightened failure risks as they strive to reach a 5% increase in GDP per capita at purchasing power parity (PIBpcPPP) within 120 months. This emphasizes the challenges faced by nations in sustaining economic development and underscores the crucial role of strategic decision-making.

Understanding time-to-achievement patterns and associated risks can assist stakeholders in evaluating economic development opportunities, designing support mechanisms, and formulating policies to enhance national resilience. The precise estimates obtained from the analysis provide valuable insights for countries seeking to optimize strategies and mitigate potential challenges on the path to economic prosperity.

*4.1 Model comparison*

Utilizing a collection of pertinent characteristics, the article evaluated the efficacy of several machine learning survival models in forecasting startup failures. For machine learning design, this approach partitioned the dataset into two subsets: a training set and a testing set. 70 percent of the rows from the data frame *df* are selected at random and assigned to the *data.train* variable. The train index variable is responsible for storing the row indices in numeric format for *data.train*. 30 percent of the initial data set is comprised of the remaining rows, which are designated as *data.test*.

This division enables the training of a model on the designated training set and the subsequent evaluation of its performance on the testing set in order to gauge its efficacy and capacity for generalization. Comparing the prediction capability of various models required the concordance index (C-index).

**Figure 2: results from different machine learning models.**

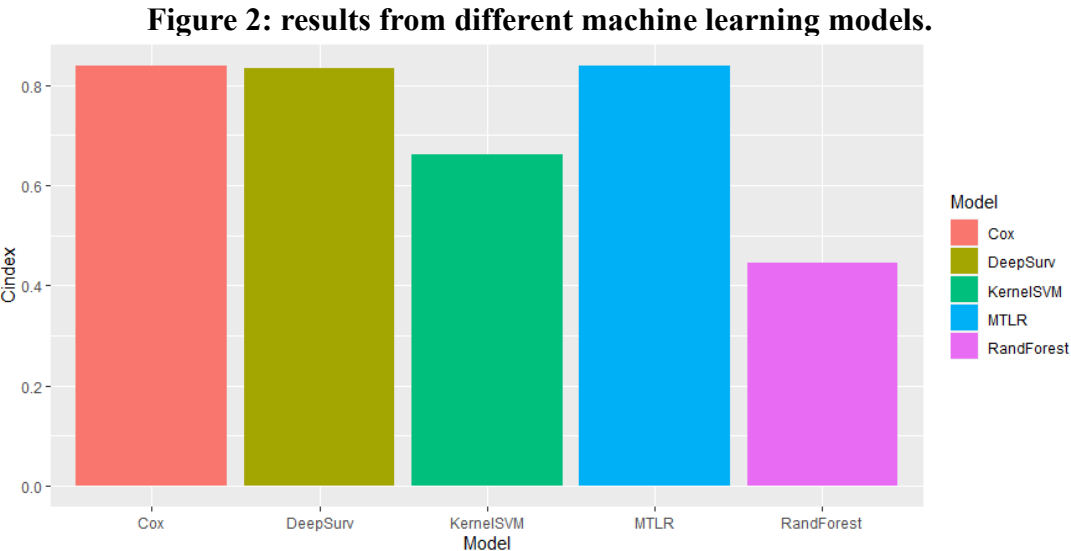

**Source**: own elaboration

The provided figure 2 furnishes a comprehensive comparative analysis of several survival models applied to a dataset focused on examining the time required for each country to attain a 5% increase in GDP per capita at purchasing power parity (PIBpcPPP) within a 120-month timeframe. This scrutiny is particularly significant in scenarios marked by a substantial number of censoring events on both the right and left sides, where advanced machine learning models often outshine conventional Cox proportional hazards models.

The Concordance Index (C-index) serves as a pivotal metric for assessing the discriminatory power of these models. Remarkably, the DeepSurv model emerges with a notable C-index of 0.833333, highlighting its adeptness in capturing intricate, non-linear relationships inherent in the dataset. This underscores its robust suitability for addressing bidirectional censoring.

Conversely, the Random Forest model demonstrates a C-index of 0.446429, signaling limitations in effectively capturing the intricacies of a small sample of censored data. While renowned for handling non-linear and complex relationships, the lower C-index suggests challenges in accurately predicting the time until a country achieves the specified 5% increase in GDP per capita.

In conjunction with these findings, both the Cox and MTLR models exhibit identical C-index values of 0.839286, showcasing their comparable performance in this analysis. Meanwhile, the KernelSVM model presents a C-index of 0.660714, indicating moderate discriminatory power in comparison to the other models.

This comparative analysis highlights that in scenarios with a limited number of observations and a low percentage of censored data, traditional models and more complex machine learning models may exhibit similar behavior. While advanced machine learning models, especially DeepSurv, demonstrate superior performance in efficiently managing datasets with censoring at both extremes under certain conditions, it is essential to acknowledge that with a small sample size and a low proportion of censored data, conventional approaches such as Cox proportional hazards, and MTLR, may demonstrate comparable effectiveness in addressing the specific challenges posed by bidirectional censoring.

*4.2 Economic perspective*

We conduct an in-depth examination of the weight matrix, shedding light on the economic implications of each variable within the context of achieving a 5% increase in GDP per capita at purchasing power parity (PIBpcPPP). Our exploration encompasses risk factors, vulnerabilities, and capabilities, providing nuanced insights crucial for informed policy formulation.

*Commercial, Financial, Endogenous, and Natural Risks:* Negative weights on commercial, financial, and natural risks, along with positive weights on endogenous risk, suggest their potential association with achieving the GDPpc target. This indicates an entrepreneurial environment where calculated risks contribute to economic growth. However, an excessive reliance on these risky strategies may amplify the potential for economic downturns or financial crises.

*Commercial Risk*: Despite consistently negative coefficients for "Commercial Risk" (-0.013113018 at time 50 and -0.001570893 at time 111), the magnitude varies. Higher commercial risk consistently decreases the probability of reaching the event.

*Financial Risk:* The consistently positive coefficient for 'Financial Risk' (0.04194862 at time 50, to 0.01840382 at time 140) indicates a potential contribution to an increased probability of reaching the event. However, it's crucial to note that other factors, such as investor perception, risk-return dynamics, and policy implications, might be at play. Further analysis and domain-specific expertise are necessary to fully understand the nuanced relationship, especially considering the negative coefficients observed for other variables in the model.

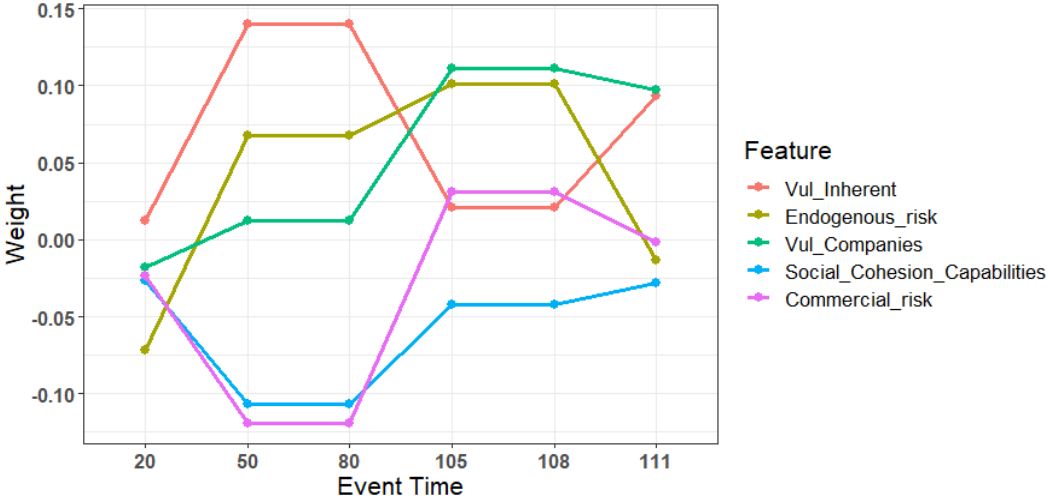

**Figure 3: Relative weight of each variable based on MTLR model.**

**Source**: own elaboration

*Endogenous Risk:* The consistently negative coefficient for 'Endogenous Risk' (-0.005939786 to -0.008532349) suggests that higher levels of endogenous risk consistently decrease the probability of reaching the specified economic event. This implies that caution is warranted when facing internal risks, as they may adversely impact the likelihood of achieving the GDPpc target over time.

*Natural Risk:* While the specific details for 'Natural Risk' are not explicitly provided in the weight matrix, its inclusion suggests a potential impact on the specified economic event. Throughout the period of time analyzed, this variable has different behavior and impacts in the analysis of the GDPpcPPP. A negative coefficient suggests a negative association with survival, while a positive coefficient indicates a positive association. A more in-depth exploration and analysis of the coefficients associated with 'Natural Risk' at different time points are necessary to comprehensively understand its role in the economic development process.

*Vulnerabilities, Inherent, Homes, and Companies:* The presence of both positive and negative weights on vulnerability variables within the weight matrix suggests nuanced implications. Countries experiencing higher inherent vulnerabilities, fragility in democratic institutions, and vulnerabilities in domestic resources, companies, and homes may find themselves either positively or negatively positioned in their pursuit of the GDPpc target.

*Inherent Vulnerability*: The consistently positive coefficient (0.07718029 to 0.01124327) indicates that inherent vulnerability consistently contributes to an increase in the probability of reaching the event.

*Vulnerability of Companies:* The consistently positive coefficients for 'Vul_Companies' across different time points (ranging from 0.0124 to 0.1109) suggest that higher vulnerability in companies is associated with an increased probability of achieving the specified economic event. This implies that addressing and mitigating vulnerabilities within the corporate sector, as indicated by 'Vul_Companies,' may play a crucial role in enhancing the likelihood of attaining the GDPpc target.

*Vulnerability of Homes:* The consistently negative coefficients for 'Vul_Homes' across different time points (ranging from -0.0676 to -0.0191) suggest that higher vulnerability in homes is associated with a decreased probability of achieving the specified economic event. This implies that strategies aimed at mitigating vulnerabilities at the household level may be crucial for improving the likelihood of attaining the GDPpc target.

*Social_Cohesion_Capabilities:* The weight matrix consistently features negative coefficients. This suggests that higher levels of social cohesion, combined with broader capabilities, are associated with a consistently decreased probability of reaching the specified GDP per capita target over the analyzed time period. The negative coefficients imply that, in this context, factors such as community resilience, inclusiveness, and social stability, encapsulated within 'Social_Cohesion_Capabilities,' may pose challenges to achieving the targeted economic outcome.

*Capabilities of the State:* The consistently positive coefficient (0.009647651) indicates that an increase in state capabilities contributes positively to the likelihood of achieving the specified economic event.

*Bias:* The positive bias in the weight matrix suggests an underlying positive force influencing the attainment of the GDPpc target, capturing unobserved factors such as favorable global economic conditions or geopolitical stability.

Based on our findings, several public policies are suggested:

*Balancing Risks and Rewards:* Countries should carefully balance the pursuit of commercial and financial opportunities with inherent risks. Calculated risk-taking is vital for economic development, but excessive exposure could lead to undesirable consequences.

*Addressing Vulnerabilities Strategically:* Policymakers should strategically address vulnerabilities, including enhancing democratic institutions and ensuring sustainable management of domestic resources, to expedite economic development.

*Investing in Capabilities:* The positive impact of state capabilities and social cohesion underscores the importance of investing in institutions and social structures. Prioritizing policies that strengthen governance, reduce corruption, and foster social harmony supports long-term economic development.

*Contextual Decision-Making:* Recognizing the context-specific nature of these weights is paramount. Tailored strategies considering the unique challenges and opportunities in each region or country are necessary for effective policy implementation.

Briefly, while the weight matrix provides valuable insights, strategic decision-making requires a nuanced understanding of the interplay between risks, vulnerabilities, capabilities, and the broader economic context. Countries aspiring to reach the GDPpc target should approach their policies with a careful balance, addressing vulnerabilities, leveraging strengths, and adapting strategies to their specific economic and social landscape.

## 5. Conclusion

The comprehensive comparative analysis presented in Figure 2 examines various survival models applied to a dataset focused on the time required for countries to achieve a 5% increase in GDP per capita at purchasing power parity (PIBpcPPP) within a 120-month timeframe. In scenarios marked by bidirectional censoring events, advanced machine learning models, particularly DeepSurv, often outshine conventional Cox proportional hazards models. The Concordance Index (C-index) serves as a pivotal metric for assessing the discriminatory power of these models.

DeepSurv emerges with a notable C-index of 0.833333, highlighting its adeptness in capturing intricate, non-linear relationships and its robust suitability for small samples scenarios. Conversely, the Random Forest model demonstrates a lower C-index of 0.446429, indicating challenges in effectively capturing censored data's intricacies. The Cox and MTLR models exhibit identical C-index values of 0.839286, showcasing comparable performance. The KernelSVM model presents a C-index of 0.660714, indicating moderate discriminatory power.

This comparative analysis underscores that, in scenarios with a limited number of observations and a low percentage of censored data, traditional models and more complex machine learning models may exhibit similar behavior. While advanced machine learning models, especially DeepSurv, demonstrate superior performance in efficiently managing datasets with censoring at both extremes under certain conditions, conventional approaches such as Cox proportional hazards and MTLR may demonstrate comparable effectiveness with a small sample size and a low proportion of censored data.

From an economic perspective, the in-depth examination of the weight matrix provides nuanced insights into risk factors, vulnerabilities, and capabilities crucial for informed policy formulation. Negative weights on commercial, financial, and natural risks, along with positive weights on endogenous risk, suggest their potential association with achieving the GDP per capita target, indicating an entrepreneurial environment where calculated risks contribute to economic growth. However, an excessive reliance on these risky strategies may amplify the potential for economic downturns or financial crises.

Addressing vulnerabilities strategically, including enhancing democratic institutions and ensuring sustainable resource management, is crucial for expediting economic development. Positive weights on vulnerability variables imply that countries facing higher inherent vulnerabilities, democratic fragility, and challenges in domestic resources, companies, and homes might find themselves better positioned to attain the GDP per capita target.

The consistently positive coefficient for inherent vulnerability indicates its consistent contribution to reaching the specified economic event. Conversely, the consistently negative coefficient for the vulnerability of companies suggests that higher vulnerability consistently decreases the probability of reaching the economic target, emphasizing the need to address and mitigate vulnerabilities in the corporate sector.

The weight matrix consistently features negative coefficients for social cohesion capabilities, suggesting that higher levels of social cohesion, combined with broader capabilities, are associated with a consistently decreased probability of reaching the specified GDP per capita target over the analyzed time period.

The positive bias in the weight matrix indicates an underlying positive force influencing GDP per capita target attainment, capturing unobserved factors such as favorable global economic conditions or geopolitical stability.

In conclusion, the weight matrix provides valuable insights, but strategic decision-making requires a nuanced understanding of the interplay between risks, vulnerabilities, capabilities, and the broader economic context. Countries aspiring to reach the GDP per capita target should approach their policies with a careful balance, addressing vulnerabilities, leveraging strengths, and adapting strategies to their specific economic and social landscape.

**Annex**

**Table 2: Country List**

| | |
|---|---|
| Antigua y Barbuda | Guatemala |
| Argentina | Guyana |
| Bahamas | Haiti |
| Barbados | Honduras |
| Belize | Jamaica |
| Bolivia | Mexico |
| Brasil | Nicaragua |
| Canada | Panama |
| Chile | Paraguay |
| Colombia | Peru |
| Costa Rica | Republica Dominicana |
| Dominica | Saint Kitts y Nevis |
| Ecuador | San Vicente y las Granadinas |
| El Salvador | Santa Lucia |
| Estados Unidos de America | Surinam |
| Grenada | Trinidad y Tobogo |
| | Uruguay |

**Figure 4: MVI and its dependent variables**

```
Call:
lm(formula = MVI ~ ., data = df[, -c(1:2)])

Residuals:
      Min        1Q    Median        3Q       Max
-0.0302975 -0.0092157 -0.0004935 0.0108401 0.0286271

Coefficients:
                            Estimate Std. Error t value Pr(>|t|)
(Intercept)                 0.169364   0.063444   2.669 0.015631 *
Natural_risk                0.126599   0.016991   7.451 6.65e-07 ***
Commercial_risk            -0.047992   0.047499  -1.010 0.325703
Financial_risk              0.070071   0.021297   3.290 0.004069 **
Endogenous_risk             0.047028   0.049875   0.943 0.358206
Vul_Inherent                0.000374   0.050679   0.007 0.994192
Vul_Companies               0.128304   0.038567   3.327 0.003753 **
Vul_Homes                   0.157584   0.034364   4.586 0.000229 ***
Capabilities_State          0.004730   0.055431   0.085 0.932943
Social_Cohesion_Capabilities 0.167742  0.032396   5.178 6.33e-05 ***
---
Signif. codes:  0 '***' 0.001 '**' 0.01 '*' 0.05 '.' 0.1 ' ' 1

s: 0.01837 on 18 degrees of freedom
  (5 observations deleted due to missingness)
Multiple R-squared: 0.9617,
Adjusted R-squared: 0.9425
F-statistic: 50.21 on 9 and 18 DF,  p-value: 6.368e-11
```
**Source**: own elaboration

**Data Sources**

https://data.imf.org/?sk=388dfa60-1d26-4ade-b505-a05a558d9a42&sId=1479329334655
https://www.statista.com/statistics/1086634/emerging-markets-bond-index-spread-latin-america-country/
https://www.theguardian.com/news/datablog/2010/apr/30/credit-ratings-country-fitch-moodys-standard#data
https://data.worldbank.org/indicator/NY.GDP.MKTP.KD.ZG
https://datos.bancomundial.org/indicator/BN.CAB.XOKA.CD
https://data.worldbank.org/indicator/FP.CPI.TOTL.ZG
http://www.worldbank.org/en/research/brief/fiscal-space
https://www.un.org/development/desa/dpad/least-developed-country-category/ldc-data-retrieval.html
https://worldjusticeproject.org/
https://www.visionofhumanity.org/public-release-data/
https://www.visionofhumanity.org/maps/#/
https://www.v-dem.net/data/the-v-dem-dataset/
https://www.v-dem.net/
https://data.worldbank.org/indicator/GC.TAX.TOTL.GD.ZS
https://data.worldbank.org/indicator/SL.UEM.TOTL.ZS
https://www.worldbank.org/en/topic/poverty
https://drmkc.jrc.ec.europa.eu/inform-index
https://www.worldbank.org/en/programs/business-enabling-environment/doing-business-legacy
https://www.hofstede-insights.com/models/national-culture/